\documentclass[10pt,tightenlines,aps,prd,twocolumn,showpacs,preprintnumbers,amsmath,amssymb]{revtex4}
\usepackage{graphicx,color}
\definecolor{darkblue}{rgb}{0,0,0.7}
\definecolor{darkred}{rgb}{0.7,0,0}
\usepackage[dvips, colorlinks, citecolor=darkblue, linkcolor=darkred, urlcolor=blue]{hyperref}
\allowdisplaybreaks
\begin{document}

\title{Displacement-noise-free gravitational-wave detection\\
with two Fabry-Perot cavities}
\author{Andrey A. Rakhubovsky and Sergey P. Vyatchanin}
\affiliation{Faculty of Physics, Moscow State University, Moscow,
119992, Russia} \email{svyatchanin@phys.msu.ru}
\date{\today}

\begin{abstract}
We propose  two detuned Fabry-Perot cavities, each pumped through both the mirrors, positioned in line as \textit{a toy model} of the gravitational-wave (GW) detector free from displacement noise of the test masses. It is demonstrated that the noise of cavity mirrors  can be completely excluded in a proper linear combination of the cavities output signals. This model is  illustrated by a simplified round trip  model (without Fabry-Perot cavities). We show that in low-frequency region the obtained displacement-noise-free response signal is  stronger than the one of the interferometer recently proposed by S.~Kawamura and Y.~Chen.
\end{abstract}

\maketitle

\section{Introduction}\label{sec_intro}
Currently the search for gravitational radiation from astrophysical
sources is conducted with the first-generation Earth-based laser
interferometers \cite{1972_EMW_detector,2000_GW_detection} (LIGO in
USA \cite{1992_LIGO,2006_LIGO_status,website_LIGO}, VIRGO in Italy
\cite{2006_VIRGO_status,website_VIRGO}, GEO-600 in Germany
\cite{2006_GEO-600_status,website_GEO-600}, TAMA-300 in Japan
\cite{2005_TAMA-300_status,website_TAMA-300} and ACIGA in Australia
\cite{2006_ACIGA_status,website_ACIGA}). The development of the
second-generation GW detectors (Advanced LIGO in USA
\cite{2002_Adv_LIGO_config,website_Adv_LIGO}, LCGT in Japan
\cite{2006_LCGT_status}) is underway.

The ultimate sensitivity of laser gravitational detectors (as belonging to class of position meters) is restricted by the Standard Quantum Limit (SQL) --- a specific sensitivity level where the measurement noise of the meter (photon shot noise) is equal to its back-action (radiation pressure noise) \cite{1968_SQL,1975_SQL,1977_SQL,1992_quant_meas}.
The sensitivity of GW detectors is also limited by a
great amount of displacements noises of various nature: seismic and
gravity-gradient noise at low frequencies (below $\sim 50$ Hz),
thermal noise in suspensions, bulks and coatings of the mirrors
($\sim 50\div 500$ Hz).

Recently S. Kawamura and Y. Chen proclaimed idea of  so called displacement-noise-free
interferometer (DFI) which are free from displacement noise of the test masses as well as from optical laser noise \cite{2004_DNF_GW_detection, 2006_DTNF_GW_detection,2006_interferometers_DNF_GW_detection}.
The most attractive feature of DFI is the straightforward overcoming of the SQL (since radiation pressure noise is canceled) without the need of implementation of very complicated and vulnerable schemes for Quantum-Non-Demolition (QND) measurements \cite{1981_squeezed_light,1982_squeezed_light,2002_conversion,1996_QND_toys_tools}.

The possibility  of GW signal separation from displacement noise of the test masses  is based on the the {\em distributed} interaction of GWs with a laser interferometer  in contrast with localized influence of mirrors positions on the light wave only in the moments of reflection. The ``payment'' for such separation is decrease of GW response. In particular, the analysis presented in \cite{2006_interferometers_DNF_GW_detection} showed that the sensitivity to GWs at low frequencies (so called long wave length approximation $L/\lambda_\text{gw}\ll 1$ or $\Omega_{\textrm{gw}}\tau\ll 1$ ($\tau =L/c$ it time of light trip between test masses separated by distance $L$, $\lambda_\text{gw}$ is wave length and $\Omega_\text{gw}$ is mean frequency of gravitational wave) turns out to be limited by the $(\Omega_{\textrm{gw}}\tau)^2$-factor for 3D (space-based) configurations and $(\Omega_{\textrm{gw}}\tau)^3$-factor for 2D (ground-based) configurations. For the signals around $\Omega_{\textrm{gw}}/2\pi\approx 100$ Hz and $L\approx 4$ km ($\tau\simeq 10^{-5}$~s), the DFI sensitivity of the ground-based detector is $(\Omega_\text{gw}\tau)^3\simeq 10^{-6}$ times
worse than the one of the conventional Michelson interferometer (i.e. a single round-trip detector). The proposed MZ-based configurations could be modified with power- and signal-recycling mirrors, artificial time-delay devices \cite{2007_time_delay}, but nevertheless, the potentially achievable sensitivity remains
incomparable with conventional non-DFI detectors.

Another approach to the displacement noise cancellation was presented in \cite{2008_toy} where
a single detuned Fabry-Perot (FP) cavity pumped through both of its movable,
partially transparent mirrors was analyzed.

In this paper we investigate model originated from a simple toy model \cite{2008_toy} of the GW detector. Our model consists of two double pumped Fabry-Perot cavities positioned in line. Each cavity is pumped through both  partially transparent mirrors.   By properly combining the signals of  output ports of the cavity an experimenter can remove the information about the fluctuations of the mirrors displacements and laser noise from the data.  The ``payment'' for isolation of the GW signal from displacement noise in our case is the suppression  of sensitivity by factor of $(\Omega_{\textrm{gw}}\tau)^2$ (resonance gain partially compensates it) as compared with conventional interferometers --- it is larger than limiting factor $(\Omega_{\textrm{gw}}L/c)^3$ of the double Mach-Zehnder 2D configuration \cite{2006_interferometers_DNF_GW_detection}. The additional advantage of proposed scheme is the possibility to use amplitude detectors instead of more complicated homodyne ones.

This paper is organized as follows. In Sec.~\ref{simple} we analyzed simplified round trip model (without any Fabry-Perot cavities). In Sec. \ref{sec_FP_cavity} we derive the response signals of a single double pumped Fabry-Perot cavity to a gravitational wave of arbitrary frequency and introduce their proper linear combination which cancels the fluctuating displacements of one of the mirrors. In Sec. \ref{sec_2FP_cavities} we analyze configuration of two  double-pumped Fabry-Perot cavities. Finally in Sec. \ref{conclusion} we discuss the physical meaning of the obtained results and briefly outline the further prospects.

\section{Simplified round trip model}\label{simple}

\begin{figure}[h,t]
\includegraphics[width=0.35\textwidth]{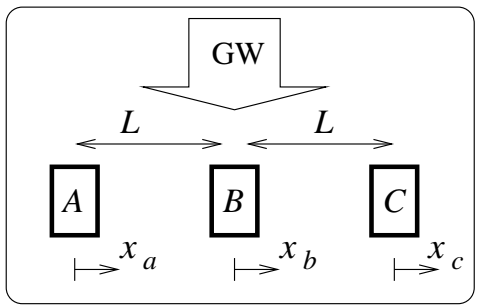}
\caption{Simplified model of displacement noise-free detector. On each platforms we place laser, detectors and reflecting mirrors. Mean distances between neighboring platforms are equal to $L$. GW propagates perpendicularly to line consisting of three platform.}\label{DL3}
\end{figure}

For clear demonstration we start from analysis of the simplest toy model \cite{2004_DNF_GW_detection} consisting of  $3$ platforms $A$, $B$ and $C$ positioned in line as shown on Fig.~\ref{DL3}. GW propagates perpendicularly to this line. We assume that lasers, detectors and mirrors are rigidly mounted on each platform which, in turn, can move as a free masses.

We restrict ourselves to the case when radiation emitted from the laser on some platform  is registered (after reflection) by detector on the same platform --- so called round trip configuration.  Actually detectors are homodyne detectors measuring the phase of incident wave.
 
Strictly speaking, in order to describe detection of light wave we have to work in the reference frame of detector, i.e. in accelerated frame. However, in our model detector is mounted on the same platform as laser which radiation detector registers and we can work in inertial laboratory frame as it was demonstrated in  \cite{2008_toy,2008_Tarabrin}. Moreover, in this case of round trip configuration we can use transverse-traceless (TT) gauge  considering GW action as effective modulation of refractive index $(1+h(t)/2)$ by weak GW perturbation metric $h(t)$. It is worth noting that in the opposite case, when laser and detector are mounted on different platforms, we should use the local Lorentz (LL) gauge --- see details in \cite{2008_Tarabrin}.

We denote the phase of the wave  emitted, for example, from platform $A$, reflected on platform $C$ and detected on platform $A$ as $\phi_{aca}$ and so on. We will not take into account fluctuations of radiations emitted by laser paying attention only on displacement noise and GW signal.

Let us measure  phase $\phi_{aba}$ (of the wave emitted from  and detected on platform $A$ after reflection from platform $B$) and phase $\phi_{bab}$ (see notations on Fig.~\ref{DL3})
\begin{align}
 \phi_{aba}(t) &= \psi_h(t)+   k\big[2x_b(t-\tau) -x_a(t) - x_a(t-2\tau)\big],\nonumber\\
 \phi_{bab}(t) &= \psi_h(t)+   k\big[-2x_a(t-\tau) +x_b(t) + x_b(t-2\tau)\big],\nonumber\\
 \label{psi}
 \psi_h (t) & \equiv\frac{\omega_0}{2}\int_{t-2\tau}^{t}h(t')dt',
\end{align}
Here $k=\omega_0/c$ is the wave vector of light emitted by laser, $\tau=L/c$ is bouncing time and $h(t)$ is perturbation of dimensionless metric originated by GW, $c$ is the speed of light.

Obviously, we can exclude information on displacement $x_a$ of platform $A$ in the following combination $\tilde C_1$:
\begin{align}
 \tilde C_1 (t)&= 2\phi_{aba} - \phi_{bab}(t+\tau)-\phi_{bab}(t-\tau)=\nonumber\\
 \label{CC1}
   &= 2\psi_h(t) - \psi_h(t+\tau) -\psi_h(t-\tau)+ \\ 
   &\qquad +k\big[2x_b(t-\tau) -x_b(t+\tau)- x_b(t-3\tau)\big].\nonumber
\end{align}
Exclusion of information on displacements of platforms $A$ in $\tilde C_1$ means that we effectively convert platform $A$  into ideal (i.e. displacement noise free) test masses for GW detection.
 
By similar way measuring phases $\phi_{bcb}$ and $\phi_{cbc}$
\begin{align}
 \phi_{cbc}(t) &= \psi_h(t)+   k\big[-2x_b(t-\tau) +x_c(t) + x_c(t-2\tau)\big],\nonumber\\
 \phi_{bcb}(t) &= \psi_h(t)+   k\big[2x_c(t-\tau) -x_b(t) - x_b(t-2\tau)\big],\nonumber
\end{align}
we can exclude information on displacement $x_c$ of platform $C$ in combination $\tilde C_2$:
 \begin{align}
  \tilde C_2 (t) &= 2\phi_{cbc} (t) -\phi_{bcb} (t-\tau)-\phi_{bcb} (t+\tau)=\nonumber\\
  \label{CC2}
    &=2\psi_h(t)-\psi_h(t-\tau) - \psi_h(t+\tau)+\\
 &\qquad +  k\big[-2x_b(t-\tau) +x_b(t+\tau) + x_b(t-3\tau)\big]\nonumber
 \end{align}

Comparing (\ref{CC1}) and (\ref{CC2}) we see that position $x_b$ makes contributions into $\tilde C_1$ and $\tilde C_2$ with opposite signs --- in contrast to the GW signal. So we should just sum $\tilde C_1$ and $\tilde C_2$ in order to exclude {\em completely} information on positions of all platforms:
\begin{align}
 \tilde C_3 (t) &= \frac{\tilde C_1(t)+\tilde C_2(t)}{2}= 2\psi_h(t)-\psi_h(t+\tau)-\psi_h(t-\tau)
\end{align}
It is useful to rewrite this formula in frequency domain:
\begin{align}
\psi_h (\Omega)&= \omega_0\tau h(\Omega)\, e^{i\Omega\tau} \, \frac{\sin\Omega\tau}{\Omega\tau}\\
\label{tildeC3}
 \tilde C_3(\Omega) &=-\omega_0\tau\, h(\Omega)\, \left(1 - e^{i\Omega\tau}\right)^2
    \left(\frac{\sin\Omega\tau}{\Omega\tau}\right)
\end{align}
In long  wave approximation ($\Omega\tau \ll 1$) we have
\begin{align}
 \tilde C_3 (t)&\simeq -\omega_0\tau^3\, \ddot h(t),\\
 \label{tildeC3app}
 \tilde C_3(\Omega) &\simeq \omega_0\tau\, (\Omega\tau)^2\, h(\Omega)
\end{align}

We see that in our simplest model the payment for separation of GW signal from displacement noise is decrease of GW response, which in long wave approximation is about $(\Omega\tau)^2$.

\section{Response of double pumped Fabry-Perot cavity to a gravitational wave}\label{sec_FP_cavity}

\begin{figure}
\includegraphics[width=0.5\textwidth]{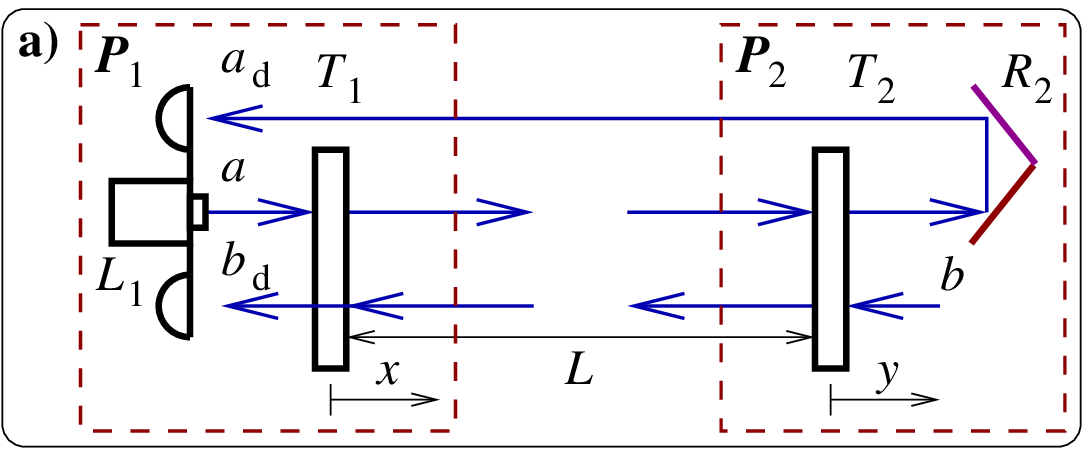}
\includegraphics[width=0.5\textwidth]{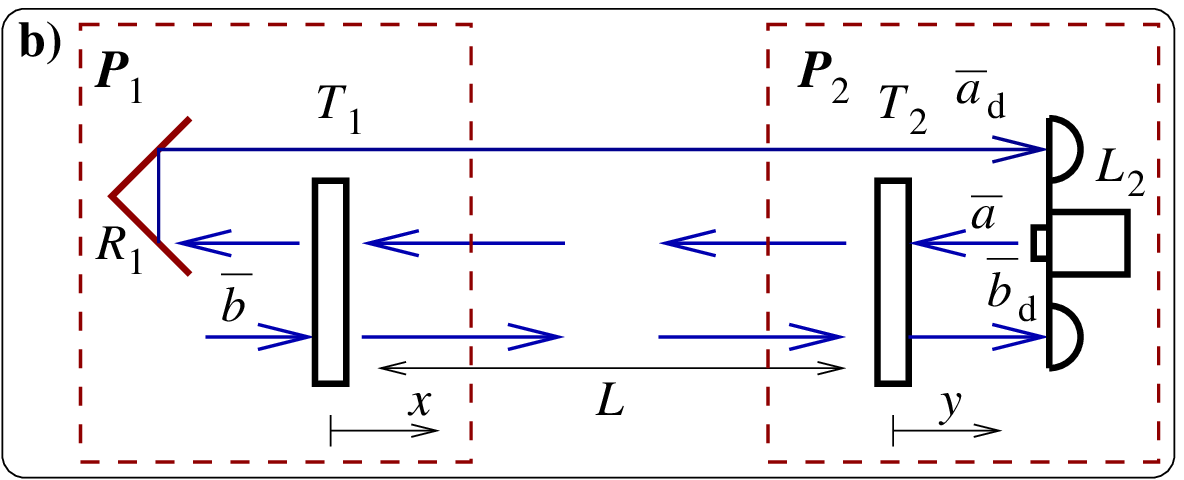}
\caption{Emission-detection scheme of {\em one} double pumped FP cavity. a) pump by laser $L_1$ through the left port is shown only. Pump laser with both detectors and input mirror are assumed to be rigidly mounted on moveable platform $\textrm{P}_1$. Transmitted wave is redirected by additional mirror $R_2$ to platform $P_1$. Transmitted and reflected wave are detected by detectors on platform $P_1$. End and additional mirror $R_2$ are assumed to be rigidly mounted on movable platform $\textrm{P}_2$. b) pump by laser $L_2$ through the right port of the same cavity with its detectors and redirecting mirror $R_1$ is shown.}
\label{T1T2oneFP}
\end{figure}

Now we can analyze model with two Fabry-Perot cavities. We start from  single double pumped FP cavity presented on Fig.~\ref{T1T2oneFP}. Pump waves in different input ports are assumed to be orthogonally polarized in order the corresponding output waves to be separately detectable and to exclude nonlinear coupling of the corresponding intracavity waves. To simplify our  model we assume that mirrors and lasers with detectors of each cavity are rigidly mounted on two movable platform (see Fig.~\ref{T1T2oneFP}), in contrast to scheme analyzed in \cite{2008_toy} with four platforms. Laser $L_1$  with its detectors and mirror with amplitude transmittance $T_1$ are rigidly mounted on movable platform $P_1$. In other words, we assume that all the elements on the platform do not move with respect to each other. Laser $L_1$ pumps the cavity from the left and we assume that the wave transmitted through the cavity is redirected to platform $P_1$ by reflecting mirror $R_2$ as shown on Fig~\ref{T1T2oneFP}a. So  waves,  emitted by this laser, are finally registered by detectors positioned on the same platform as laser. The  mirror with amplitude transmittance $T_2$ and laser $L_2$ pumping cavity from the right with his detectors are rigidly mounted on platform $P_2$.  We assume that amplitude transmission coefficients of mirrors are small: $T_1,\ T_2\ll 1$. We put mean distance between the mirrors to be equal to $L$. Without the loss of generality we assume the cavity to be lying in the plane perpendicular to direction of GW and  along one of the GW principal axes.

It is convenient to represent the electric field operator of the light wave as a sum of (i) the ``strong'' (classical) plane monochromatic wave (which approximates the light beam with cross-section $S$) with
amplitude $A$ and frequency $\omega_0$ and (ii) the ``weak'' wave describing quantum fluctuations of the electromagnetic field:
\begin{subequations}
\label{E}
\begin{align}
    E(x,t)&=\sqrt{\frac{2\pi\hbar\omega_0}{Sc}}\,
    \Bigl[A+a(x,t)\Bigr]e^{-i(\omega_0t\mp k_0x)}+{\textrm{h.c.}},\\
    a(x,t)&=\int_{-\infty}^{+\infty}
    a(\omega_0+\Omega)e^{-i\Omega\left(t\mp x/c\right)}\,\frac{d\Omega}{2\pi},\nonumber
\end{align}
\end{subequations}
with amplitude $a(\omega_0+\Omega)$ (Heisenberg operator to be
strict) obeying the commutation relations:
\begin{align*}
    \bigl[a(\omega_0+\Omega),a(\omega_0+\Omega')\bigr]&=0,\\
    \bigl[a(\omega_0+\Omega),a^+(\omega_0+\Omega')\bigr]&=2\pi\delta(\Omega-\Omega').
\end{align*}
This notation for quantum fluctuations $a(\omega_0+\Omega)$ is convenient since it coincides exactly with the Fourier representation of the classical fields. For briefness throughout the paper we denote
\begin{equation}
a\equiv a(\omega_0+\Omega),\quad a^+_-\equiv a^+(\omega_0-\Omega) \nonumber
\end{equation}
and we omit the $\sqrt{2\pi\hbar\omega_0/Sc}$-multiplier. For convenience throughout the paper we denote mean amplitudes by block letters and corresponding small additions by {\em the same} small letter as in (\ref{E}). We assume that input laser fields are in coherent state (it means that fluctuational amplitude $a(\omega_0+\Omega)$ describes vacuum fluctuations) and fields on the non-pumped port are in vacuum state.

In our model, as in simplified model analyzed in previous section, detectors are mounted on the same platform as laser which radiation detectors register and we can work in inertial laboratory frame   \cite{2008_toy,2008_Tarabrin} considering GW action as effective modulation of refractive index $(1+h(\Omega)/2)$ by weak GW perturbation metric $h(\Omega)$.

First, we consider pump by laser $L_1$ shown in the Fig.~\ref{T1T2oneFP}a. Using calculations presented in Appendix \ref{derivation} we can write down formulas for small complex amplitudes $a_d,\ b_d$ of waves detected on platform $P_1$ (see notations on Fig.~\ref{T1T2oneFP}a):
\begin{align}
\label{ad}
a_d&=-\theta_0\psi\big[{\cal T} a+ {\cal R}_2 b \big]+\\
   & + \frac{iT_2\vartheta_0^2}{1-R_1R_2\vartheta_0^2\psi^2}
        \left(\frac{1 +\psi^2}{2}u_x-\psi (u_y+u_h)\right),\nonumber\\
\label{bd}
b_d&={\cal R}_1 a+ {\cal T} b +\\
&   +     \frac{iT_1R_2\vartheta_0^2}{1-R_1R_2\vartheta_0^2\psi^2}
   \left( \frac{1+\psi^2}{2} u_x- \psi (u_y+u_h)\right),\nonumber\\
&\text{where}\quad  \psi = e^{i\Omega\tau},\quad \theta_0=e^{i\delta\tau},\quad \tau=\frac{L}{c}\, .
\end{align}
Here $\delta$ is detuning between laser frequency and resonance frequency of cavity, $R_1=\sqrt{1-T_1^2}$, $R_2=\sqrt{1-T_2^2}$ are reflectivities of mirrors, by calligraph letters we denote coefficients of cavity's transparency and reflectivities:
\begin{align}
 {\cal T} & =\frac{-\vartheta_0\psi T_1T_2}{1-R_1R_2\vartheta_0^2\psi^2},\\
 {\cal R}_1 &= \frac{R_2\vartheta_0^2\psi^2-R_1}{1-R_1R_2\vartheta_0^2\psi^2},\quad
 {\cal R}_2 = \frac{R_1\vartheta_0^2\psi^2-R_2}{1-R_1R_2\vartheta_0^2\psi^2}
\end{align}
The influence of fluctuational (non-geodesic) displacements $x,\ y$ in (\ref{ad}, \ref{bd}) (to be strict its Fourier representations) is described by values $u_x, \ u_y$:
\begin{align}
 u_x &= A_{in} \, 2ik x(\Omega), \quad u_y=A_{in}\, 2iky(\Omega),\\
 \label{Ain}
  A_{in}&=\frac{iT_1A}{1-R_1R_2\vartheta_0^2},\quad
    k=\frac{\omega_0}{c},
\end{align}
where $A_{in}$ is mean amplitude of wave circulating inside the cavity, we assume $A_{in}$ to be real (see also Fig.~\ref{fpthroughT1T2fourP} in Appendix~\ref{derivation}), $A$ is mean amplitude of wave emitted by laser $L_1$ (to be strict amplitude of wave falling on mirror with transparency $T_1$). Interaction of light with GW in (\ref{ad}, \ref{bd}) is described by dimensionless metric perturbation $h$ through value $u_h$:
\begin{align}
 u_h& = A_{in}\, ikL\,h(\Omega)\, \frac{\sin \Omega\tau}{\Omega\tau}\, .
\end{align}

It is worth emphasizing that in this scheme  we can analyze amplitude quadrature of the wave falling on detector which can be measured by simple amplitude detector, but not, for example, phase quadrature which requires for registration more complicated homodyne detector. For amplitude component we have (see details in Appendix~\ref{derivation})
\begin{align}
 a_d^{(a)} & =
     \frac{\psi\big(\vartheta_0{\cal T}^*_-a^+_- -\vartheta_0^*{\cal T}a\big) +
        \psi\big(\vartheta_0{\cal R}_{2-}^* b_{-}^+ - \vartheta_0^*{\cal R}_2 b\big)}{i\sqrt 2}+\nonumber\\
 \label{adAmp}
    & + \frac{T_2R_1R_2\psi^2\big(\vartheta_0^2-(\vartheta_0^*)^2\big)}{
    \sqrt 2\, \big(1-R_1R_2\vartheta_0^2\psi^2\big)\big(1-R_1R_2(\vartheta_0^*)^2\psi^2\big)}\times\\
    &\qquad \times \left(\frac{\big(1+\psi^2\big)}{2}\, u_x -\psi (u_y+u_h)\right),\nonumber\\
\label{bdAmp}
 b_d^{(a)} & =  \frac{\big(R_2(\vartheta_0^*)^2-R_1\big)\big({\cal R}_1 a +{\cal T}b_2\big)
    }{i|R_2\vartheta_0^2 - R_1|\sqrt 2}+\big\{\text{h.c.}\big\}_- \\
  +&  \frac{T_1R_1R_2\big(\vartheta_0^2 -(\vartheta_0^*)^2\big)\big(R_2^2\psi^2-1\big)
    }{\sqrt 2|R_2\vartheta_0^2-R_1|
        \big(1-R_1R_2\vartheta_0^2\psi^2\big)\big(1-R_1R_2(\vartheta^*_0)^2\psi^2\big)}\times \nonumber\\
  &\qquad \times  \left(\frac{\big(1+\psi^2\big)}{2}\, u_x -\psi (u_y+u_h)\right).\nonumber
\end{align}
Here notation \{h.c.\}$_-$ means hermitian conjugate with replacing $\Omega\to -\Omega$.

We see that  platform displacements ($u_x,\ u_y$) contributes to output amplitude quadratures $a_d^{(a)},\ b_d^{(a)}$ in the same combination. Hence, we can make such combinations of them which will be  free from technical laser noise (it means exclusion of terms proportional to $a$ and $a^+_-$ in (\ref{adAmp}, \ref{bdAmp})). If so, only fundamental quantum noise will limit sensitivity. However, additional detailed analysis is required because  such exclusion of technical laser noise may be possible only partially (for example, only in finite bandwidth).

As platform displacements make the same contributions into output amplitude quadratures $a_d^{(a)},\ b_d^{(a)}$ we can use any of them. Below  we will use for manipulation only amplitude quadrature $a_d^{(a)}$ writing it as following
\begin{align}
 a_d^{(a)} & = a_{fl}+\nonumber\\
 \label{adAmp2}
    & + \frac{T_2R_1R_2\psi^2\big(\vartheta_0^2-(\vartheta_0^*)^2\big)}{
    \sqrt 2\, \big(1-R_1R_2\vartheta_0^2\psi^2\big)\big(1-R_1R_2(\vartheta_0^*)^2\psi^2\big)}\times\\
    &\qquad \times \left(\frac{\big(1+\psi^2\big)}{2}\, u_x -\psi (u_y+u_h)\right),\nonumber
\end{align}
and keeping in mind that fluctuational component $a_{fl}$ (it is terms proportional $a,\ a^+_-,\ b,\ b^+_-$ in (\ref{adAmp})) mainly corresponds to zero field fluctuations and technical laser noise is  excluded, at least partially.

Now we can write down formula for output field pumping by laser $L_2$  from opposite port (see Fig.~\ref{T1T2oneFP}b). We assume that radiation from laser $L_2$ is polarized normally to radiation emitted by laser $L_1$. We denote all values by the same letters as above but mark them by bar  $\bar{}$. For simplicity we assume that excited by laser $L_2$ mean amplitude $\bar A_{in}$ of the wave circulating inside the cavity is equal to $A_{in}$: $\bar A_{in}=A_{in}$. Also for simplicity we  assume that detuning of laser $L_2$ is the same as detuning of laser $L_1$, i.e. $\bar\vartheta_0=\vartheta_0$. Then by using the following substitutions:
\begin{align*}
 T_{1,2}&\to T_{2,1},\quad R_{1,2}\to R_{2,1}, \\
   u_x & \to - u_y,\quad u_y \to -u_x,\quad u_h\to u_h
\end{align*}
we rewrite formula (\ref{adAmp2}) for amplitude quadrature $\bar a_d^{(a)}$
\begin{align}
 \bar a_d^{(a)} & = \bar a_{fl}+\nonumber\\
 \label{adAmp3}
    & + \frac{T_1R_1R_2\psi^2\big( \vartheta_0^2-(\vartheta_0^*)^2\big)}{
    \sqrt 2\, \big(1-R_1R_2 \vartheta_0^2\psi^2\big)
    \big(1-R_1R_2( \vartheta_0^*)^2\psi^2\big)}\times\\
    &\qquad \times \left(\frac{\big(1+\psi^2\big)}{2} (-u_y) -\psi (-u_x+u_h)\right).\nonumber
\end{align}

Comparing formulas (\ref{adAmp2}) and (\ref{adAmp3}) we see that platform displacements ($u_x$ and $u_y$) make different contributions into the output quadrature. It allows to exclude one of these displacements.

\section{Displacements exclusion in configuration of two double pumped Fabry-Perot cavities}\label{sec_2FP_cavities}

Analyzing formulas (\ref{adAmp2}, \ref{adAmp3}) we find that one can exclude information on $u_y$ in the following combination:
\begin{align}
 C_1&=\sqrt \frac{T_1}{T_2}\, \frac{1+\psi^2}{2}\, a_d^{(a)} -
        \sqrt \frac{T_2}{T_1}\, \psi\, \bar a_d^{(a)}=\nonumber\\
\label{C1}
 & =    \sqrt \frac{T_1}{T_2}\, \frac{1+\psi^2}{2}\, a_{fl} -
        \sqrt \frac{T_2}{T_1}\, \psi\, \bar a_{fl}+\\
 & \quad + \frac{\sqrt{T_1T_2}\,R_1R_2\psi^2\big(\vartheta_0^2-(\vartheta_0^*)^2\big)}{
    \sqrt 2\, \big(1-R_1R_2\vartheta_0^2\psi^2\big)\big(1-R_1R_2(\vartheta_0^*)^2\psi^2\big)}
    \times\nonumber\\
    &\qquad \times \left(\left[\frac{1 -\psi^2}{2}\right]^2 u_x -\psi\,\frac{(1-\psi)^2}{2}\, u_h\right)
    . \nonumber
\end{align}

It is a very important result --- exclusion of information on $u_y$ is equivalent to conversion of platform $P_2$ into ideal mass, which is free from fluctuational displacement.  The payment for such conversion is decrease of GW rresponseby factor approximately $\sim (1-\psi)^2$ (it is about $\sim(\Omega \tau)^2$ in long wave approximation) as compared with conventional laser GW detector.

\begin{figure}
\includegraphics[width=0.5\textwidth]{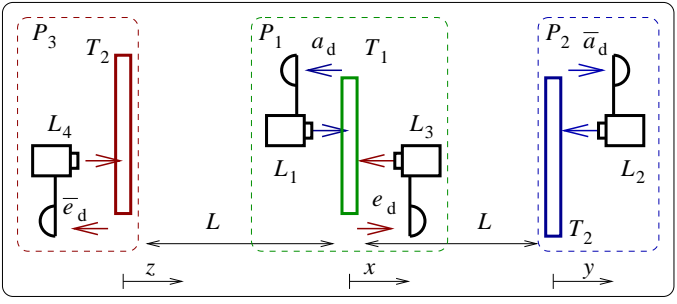}
\caption{Configuration of two doubled pumped Fabry-Perot cavities. The right Fabry-Perot cavity is the same as shown on Fig,~\ref{T1T2oneFP}, the redirecting mirrors and detectors registering reflected waves $b_d,\ \bar b_d$ are not shown. The left Fabry-Perot cavity is identical to right cavity having the common mirror with transparency $T_1$ rigidly mounted on platform $P_1$. Left cavity is pumped by lasers $L_3$ and $L_4$ and we detect transmitted waves $e_d,\ \bar e_d$ redirected by additional mirrors. These mirrors (as well as detector s for registration of reflected from cavity waves) are not shown. }
\label{T1T2threeP}
\end{figure}

Now we have to exclude information on $u_x$ (i.e. displacement $x$ of platform $P_1$). It can be done in configuration of two double pumped Fabry-Perot cavities. Let us add  second Fabry-Perot cavity (left cavity on Fig.~\ref{T1T2threeP}) positioned in line with first cavity considered above. For simplicity we assume that parameters of both cavities are identical and that amplitudes and detunings of lasers $L_3,\ L_4$ pumped second cavity are the same as of lasers $L_1,\ L_2$ correspondingly. The mirror with transparency $T_1$ is common for two cavities. This mirror and laser $L_3$ with its detectors are rigidly mounted on the same platform $P_1$ as well as laser $L_1$ with its detectors. The other mirror of second cavity and laser $L_4$ with its detectors are rigidly mounted on platform $P_3$, we denote its position  by $z$.

In order to calculate formulas for amplitude quadratures output waves $e_d^{(a)},\ \bar e_d^{(a)}$ of second cavity just rewriting formulas (\ref{adAmp2}, \ref{adAmp3}) for amplitude quadratures $a_d^{(a)},\ \bar a_d^{(a)}$ we apply following substitutions:
\begin{subequations}
 \label{subs}
\begin{align}
 u_y & \to -u_z,\quad u_x \to -u_x,\quad u_h \to u_h, \\
    &a_{fl}\to e_{fl}, \quad \bar a_{fl}\to \bar e_{fl}.
\end{align}
\end{subequations}
Here $e_{fl},\ \bar e_{fl}$ are corresponding fluctuational component originated by input fluctuational fields of the second cavity.

We can also exclude information on displacement $z$ in combination $C_2$ by the same way as we excluded displacement $y$ in combination $C_1$. One can write this combination $C_2$ free from displacement $z$ using substitutions (\ref{subs}):
\begin{align}
 C_2&=\sqrt \frac{T_1}{T_2}\, \frac{1+\psi^2}{2}\, e_d^{(a)} -
        \sqrt \frac{T_2}{T_1}\, \psi\, \bar e_d^{(a)}=\nonumber\\
\label{C2}
 & =    \sqrt \frac{T_1}{T_2}\, \frac{1+\psi^2}{2}\, e_{fl} -
        \sqrt \frac{T_2}{T_1}\, \psi\, \bar e_{fl}+\\
 & \quad + \frac{\sqrt{T_1T_2}\,R_1R_2\psi^2\big(\vartheta_0^2-(\vartheta_0^*)^2\big)}{
    \sqrt 2\, \big(1-R_1R_2\vartheta_0^2\psi^2\big)\big(1-R_1R_2(\vartheta_0^*)^2\psi^2\big)}
    \times\nonumber\\
    &\qquad \times \left(-\left[\frac{1 -\psi^2}{2}\right]^2 u_x -\psi\,\frac{(1-\psi)^2}{2}\, u_h\right)
    . \nonumber
\end{align}

Comparing (\ref{C1}, \ref{C2}) we see that value $u_x$ makes contributions into $C_1$ and $C_2$ with the opposite signs, whereas GW contributions (i.e. $u_h$) have the same sign (it is obvious consequence of tidal nature of GW). So in order to exclude $u_x$ we should just sum $C_1$ and $C_2$:
\begin{align}
 C_\text{DFI} &= (C_1 + C_2)/\sqrt 2 =\nonumber\\
 & =  \sqrt \frac{T_1}{T_2}\, \frac{1+\psi^2}{2}\, \frac{\big(a_{fl}+e_{fl}\big)}{\sqrt 2} -
        \sqrt \frac{T_2}{T_1}\, \psi\, \frac{\big(\bar a_{fl}+\bar e_{fl}\big)}{\sqrt 2}+ \nonumber\\
 & \quad + \frac{\sqrt{T_1T_2}\,R_1R_2\psi^2\big(\vartheta_0^2-(\vartheta_0^*)^2\big)}{
    2\, \big(1-R_1R_2\vartheta_0^2\psi^2\big)\big(1-R_1R_2(\vartheta_0^*)^2\psi^2\big)}
    \times\nonumber\\
 \label{CDFI}
    &\qquad \times \left( -\psi\,\big(1-\psi\big)^2\, u_h\right) . 
\end{align}
Comparing combination $C_\text{DFI}$ with amplitude quadrature $a^{(a)}$ (\ref{adAmp2}) we see that gravitational signal in $C_\text{DFI}$ is smaller by factor $(1-\psi)^2$ which in  approximation of long gravitational wave length $L/\lambda_\text{grav}\ll 1$ (or $\Omega\tau\ll 1$) is about $(\Omega\tau)^2$.
It is the same decrease of GW signal as in combination  $\tilde C_3$ (\ref{tildeC3}, \ref{tildeC3app})) for simplified model considered above (the only difference is the presence of resonance gain in (\ref{CDFI})).

Assuming  $T_1,\ T_2\ll 1$ and $\Omega\tau\ll 1$ we rewrite $C_\text{DFI}$ in narrow band  approximation:
\begin{align}
 C_\text{DFI} &\simeq   \sqrt \frac{T_1}{T_2}\, \frac{\big(a_{fl}+e_{fl}\big)}{\sqrt 2} -
        \sqrt \frac{T_2}{T_1}\, \frac{\big(\bar a_{fl}+\bar e_{fl}\big)}{\sqrt 2}+ \nonumber\\
 \label{CDFIapp}
   & \quad + \frac{\sqrt{T_1^3T_2}}{T_1^2+T_2^2}\,\frac{ \gamma\delta\,\Omega
  }{(\gamma-i\Omega)^2+\delta^2}\, \Omega\tau
        \, \frac{A\, kL\, h(\Omega)}{\gamma-i\delta},
\end{align}
where $\gamma=(T_1^2+T_2^2)/4\tau$ is the relaxation rate (half bandwidth) of Fabry-Perot cavity.

Recall that in a simplest detector with two test masses and only one round trip of light between them  gravitational signal is about $AkLh$ with the same value of fluctuational field. So assuming in (\ref{CDFIapp}) that $\delta\approx\gamma\approx \Omega$ and $T_1\approx T_2$ we see that signal-to-noise ratio of our cavities operating as displacement noise free detector  is smaller by factor about $\sim \Omega\tau$ as compared with simplest detector.

\section{Conclusion}\label{conclusion}

In this paper we have analyzed the operation of two Fabry-Perot cavities positioned in line, performing the displacement-noise-free gravitational-wave detection. We have demonstrated that it is possible to construct a linear combination of four response signals which cancels displacement fluctuations of the mirrors. At low frequencies the GW response of our cavities turns out to be better than that of the Mach-Zehnder-based DFIs \cite{2006_interferometers_DNF_GW_detection} due to the different mechanisms of noise cancellation.

Additional advantage of proposed DFI configuration is that one can measure {\em amplitude} quadratures of output fields which can be registered by usual amplitude detection but, for example, not phase quadrature  requiring for registration more complicated homodyne detection.

It seems that for amplitude detection of {\em transmitted} wave we can position amplitude detector on opposite platform. Then we will have no need to redirect transmitted wave back to platform with laser. For example, amplitude quadrature $a^{(a)}$ can be registered by detector positioned on platform $P_2$ --- see Fig.~\ref{T1T2oneFP}a.  However, this proposition requires additional analysis because for our consideration we used inertial frame and it is possible only if laser and detectors, registered radiation emitted by laser, are positioned on the same platform \cite{2008_toy,2008_Tarabrin}.

Due to reflected and transmitted waves carry the same information on mirrors displacement (compare formulas (\ref{adAmp}, \ref{bdAmp})) we have additional possibility to exclude, at least partially, laser technical noise (of course, fundamental quantum laser noise can not be not excluded).

We show that considered DFI with two Fabry-Perot cavities is similar to the simplest round trip configuration shown in Fig.~\ref{DL3}.

For simplicity we have analyzed three platform configuration. The configurations with larger number of movable platform is more realistic and it may provide better sensitivity. For example, the middle platform may be splitted into three platforms: two platforms  with mirrors and one platform (between them) with lasers and detectors. Variants of such configurations are under investigation now.

The proposed configuration of DFI may be a promising candidate for the future generation of GW detectors  with  displacement  and laser noise exclusion which, in turn, will allow to overcome standard quantum limit.

\acknowledgments
We would like to thank V.B. Braginsky, F.Ya. Khalili and S.P. Tarabrin for fruitful discussions. This work was supported by LIGO team from Caltech and in part by NSF and Caltech grant PHY-0651036 and by Grant of President of Russian Federation NS-5178.2006.2.

\appendix

\section{Derivation of formulas for Fabry Perot cavity}\label{derivation}

In this Appendix we derive formulas (\ref{ad}, \ref{bd}) for complex amplitudes  and (\ref{adAmp}, \ref{bdAmp}) for amplitude quadratures for single Fabry-Perot cavity pumped by laser from the left.

\begin{figure}
\includegraphics[width=0.5\textwidth]{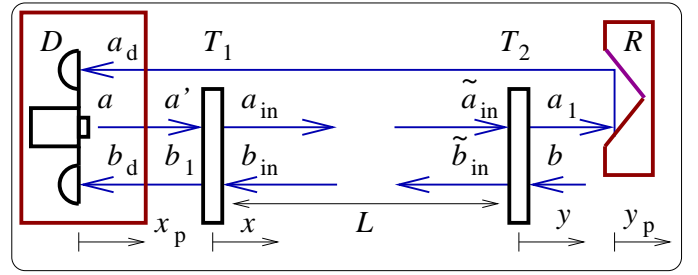}
\caption{Detailed scheme of measurement (generalization of shown on the Fig.~\ref{T1T2oneFP}a). Cavity mirrors are movable, laser and detectors are placed on detecting platform D, additional
 mirror is placed on reflecting platform $R$.}\label{fpthroughT1T2fourP}
\end{figure}

For methodical purpose we start from general case when laser with detectors, mirrors and additional mirror are mounted on separated rigid movable platform each as shown on Fig.~\ref{fpthroughT1T2fourP}. Below we use notations on Fig.~\ref{fpthroughT1T2fourP}. First we find complex mean amplitudes, writing boundary conditions on  right and left mirror:
\begin{align*}
\tilde B_{in} &= -R_2\tilde A_{in}, \quad A_1 = i\tilde A_{in} T_2,\\
A_{in}&=  iT_1\, A-R_1\, B_{in},\quad B_1=iT_1\,B_{in}-R_1\,A
\end{align*}
From these equations and obvious relations
$\tilde A_{in} = A_{in} \vartheta_0$ and $B_{in} = \tilde B_{in}  \vartheta_0$ one can find formula (\ref{Ain}) for $A_{in}$ and for mean output fields:
\begin{align}
\label{B1A2}
& B_1=\frac{A\big(R_2\vartheta_0^2-R_1\big)}{1-R_1R_2\vartheta_0^2},\quad
&  A_1 =  \frac {-T_1T_2\vartheta_0\, A}{1 -R_1R_2\vartheta_0^2}.
\end{align}

To find small amplitudes inside cavity we write down boundary condition on right and left mirrors correspondingly:
\begin{align}
\label{tildebin}
\tilde b_{in}&=    -R_2\tilde a_{in} + iT_2\,b -R_2\vartheta_0u_y\\
\label{ainA}
a_{in}&=iT_1\, a' - R_1b_{in} -R_1R_2\vartheta_0^2u_x.
\end{align}
And taking into account GW action as effective variation of refractive index $1+h/2$
\begin{align}
\tilde a_{in}&=
    \vartheta_0\psi a_{in} + A_{in}\vartheta_0\, i\,j(\Omega),\\
b_{in} &=
    \vartheta_0\psi \tilde b_{in} -R_2 A_{in}\vartheta_0^2\,i\,j(\Omega),\\
j&=\frac{\omega_0h(\Omega)}{2} \int_{t-\tau}^t e^{-i\Omega (t'-t)}\, dt'=
   \frac{kLh}{2}\left(\frac{1-e^{i\Omega\tau}}{-i\Omega\tau}\right), \nonumber\\
&\psi u_h = A_{in}\big(1+\psi\big) j\, \nonumber
\end{align}
we find small amplitudes inside cavity:
\begin{align}
\label{ainFin}
a_{in}&= \frac{iT_1\, a' -iT_2R_1\vartheta_0\psi b}{1-R_1R_2\vartheta^2\psi^2} +
   \frac{R_1R_2\vartheta_0^2\big(\psi (u_y+u_h)-u_x\big)}{1-R_1R_2\vartheta^2\psi^2},\\
\label{binFin}
b_{in} &=\frac{-iT_1R_2\vartheta_0^2\psi^2\, a' +iT_2\vartheta_0\psi b}{1-R_1R_2\vartheta_0^2\psi^2}+\\          &\qquad + \frac{R_2 \vartheta_0^2\psi\big(
   R_1R_2\vartheta_0^2\psi u_x - u_y-u_h\big)}{1-R_1R_2\vartheta_0^2\psi^2}.\nonumber
\end{align}

Now using second boundary condition on right  mirror we can find transmitted wave $a_1$:
\begin{align}
a_1 &= -R_2\, b +iT_2\, \tilde a_{in} =   \nonumber\\
\label{a1}
= & {\cal R}_2 b +{\cal T}a'
   - \frac{iR_1R_2T_2\vartheta_0^3\psi}{1-R_1R_2\vartheta_0^2\psi^2}\,
   \left( u_x-\psi (u_y +u_h)\right).
\end{align}

By the same manner from second boundary condition on left mirror we find reflected wave $b_1$
\begin{align}
b_1 &= -R_1\, a' +iT_1\, b_{in} +(-R_1 A)\, 2ikx=\nonumber \\
\label{b1}
&={\cal R}_1 a'+ {\cal T}b + \frac{i(R_1-R_2\vartheta_0^2)}{T_1}\, u_x+\\
& \qquad + \frac{iT_1R_2\vartheta_0^2}{1-R_1R_2\vartheta_0^2\psi^2}
   \big(  u_x- \psi (u_y+u_h)\big).\nonumber
\end{align}
Here we write formula for $b_1$ in this form in order to extract term  proportional to same combinations of  mirrors positions as in (\ref{a1}).

In order to express fields $a_1, \ b_1$ through small amplitude $a$ describing laser fluctuations we should substitute  in (\ref{a1}, \ref{b1})
\begin{align}
 a'& = \big(a -Aikx_p\big).
\end{align}
Now we can find field $b_d$ falling on detector
\begin{align}
 b_d &=b_1 - B_1\, ikx_p=
        {\cal R}_1 a+ {\cal T} b +
        \frac{i(R_1-R_2\vartheta_0^2)}{T_1}\, u_x+ \nonumber\\
 \label{bdfin}
 &\quad +       \frac{iT_1R_2\vartheta_0}{1-R_1R_2\vartheta^2}
   \big( \vartheta_0 u_x- \vartheta (u_y+u_h)\big) -\\
  -& \frac{A_{in} \big(1-R_1R_2\vartheta_0^2\big) ikx_p }{iT_1}\left(
    \frac{R_2\vartheta^2-R_1}{1-R_1R_2\vartheta^2}+
    \frac{R_2\vartheta_0^2-R_1}{1-R_1R_2\vartheta_0^2}\right).\nonumber
\end{align}
 
Using (\ref{a1}) for transmitted wave $a_1$ we find formula for amplitude $a_d$ falling on detector:
\begin{align}
 a_d &= \vartheta \big(-a_1 -A_1\, 2iky_p\big) +
    (-\vartheta_0 A_1) ij - (-\vartheta_0 A_1) ikx_p=\nonumber\\
&= -\vartheta_0\psi \big({\cal T} a +{\cal R}_2b\big)
    + \frac{iT_2R_1R_2\vartheta_0^4\psi^2
        \big(u_x-\psi (u_y+u_h)\big)}{1-R_1R_2\vartheta_0^2\psi^2}-\nonumber\\
\label{adfin}
&\qquad    - iT_2\vartheta_0^2\psi \big(u_h +A_{in}\, 2iky_p\big)+\\
&\qquad +     iT_2 \vartheta_0^2A_{in}ikx_p\left(1 +\psi^2
         \frac{\big(1-R_1R_2\vartheta_0^2\big)}{\big(1-R_1R_2\vartheta_0^2\psi^2\big)}
          \right).\nonumber
\end{align}
Now substituting $x_p=x$ and $y_p=y$ into (\ref{bdfin}, \ref{adfin}) one can obtain formulas (\ref{ad}, \ref{bd}).

We define amplitudes quadratures of fields falling on detectors as fallowing:
\begin{align*}
a_d^{(a)} & \equiv\frac{-A_d^*a_d-A_da_{d-}^+}{|A_d|\sqrt 2}=
    \frac{(\vartheta_0^*)^2a_d - \vartheta_0^2 a_{d-}^+}{i \sqrt 2},\\
b_d^{(a)} & \equiv \frac{B_d^*b_d+B_db_{d-}^+}{|B_d|\sqrt 2}=
    \frac{\big(R_2(\vartheta_0^*)^2-R_1\big)b_d -
        \big(R_2\vartheta_0^2-R_1\big)b^+_{d-}}{i|R_2\vartheta_0^2-R_1|\sqrt 2}\, .
\end{align*}
Substituting (\ref{ad}, \ref{bd}) into these formulas one get finally formulas (\ref{adAmp}, \ref{bdAmp}) for amplitude quadratures $a_d^{(a)},\ b_d^{(a)}$.


\end{document}